\documentclass[journal,twoside,web]{ieeecolor}
\usepackage{generic}
\usepackage{cite}
\usepackage{amsmath,amssymb,amsfonts}
\usepackage{algorithmic}
\usepackage{graphicx}
\usepackage{subfig}
\usepackage{algorithm,algorithmic}
\usepackage{hyperref}
\usepackage{textcomp}

\usepackage{multirow}

\def\BibTeX{{\rm B\kern-.05em{\sc i\kern-.025em b}\kern-.08em
    T\kern-.1667em\lower.7ex\hbox{E}\kern-.125emX}}
\begin{document}
\title{Edge-AI for Heart Murmur Screening in Low-Resource Settings}
\author{Andrea De Simone, Noemi Giordano, Silvia Seoni, Kristen M. Meiburger, and Fabrizio Riente
%\thanks{\textcolor{red}{\textbf{TO DO}} This paragraph of the first footnote will contain the date on 
%which you submitted your paper for review. It will also contain support 
%information, including sponsor and financial support acknowledgment. For 
%example, ``This work was supported in part by the U.S. Department of 
%Commerce under Grant 123456.'' }
\thanks{A. De Simone and F. Riente are with the QNANO group, Department of Electronics and Telecommunications, Politecnico di Torino, Torino, Italy (e-mail: andrea.desimone@polito.it, fabrizio.riente@polito.it). }
\thanks{N. Giordano is with the Department of Health Science and Technology, Aalborg University, Aalborg, Denmark (e-mail: nogi@hst.aau.dk).}
\thanks{S. Seoni and K.M. Meiburger are with InVisioning Lab, Polito\textsuperscript{BIO}Med Lab, Department of Electronics and Telecommunications, Politecnico di Torino, Torino, Italy.
(e-mail: silvia.seoni@polito.it, kristen.meiburger@polito.it).}}

\maketitle

\begin{abstract}
Early and reliable detection of heart murmurs is essential for the timely diagnosis of cardiovascular diseases, yet traditional auscultation remains subjective and dependent on expert interpretation. This work investigates an artificial intelligence (AI)-based approach for heart murmur detection using the CirCor Heart Sound dataset, with a particular focus on deployment in low-cost and resource-constrained embedded platforms.
A lightweight convolutional neural network (CNN) is developed using spectrogram features extracted from heart sound recordings. The complete processing pipeline, from feature extraction to final prediction, is implemented and validated on an STM32F401 microcontroller (MCU). To enable real-time execution within the limited hardware resources of the target platform, memory- and latency-oriented optimizations together with fixed-point processing are employed.
Experimental results demonstrate that accurate murmur detection can be achieved despite the severe computational constraints of the embedded device. The proposed system achieves a patient-level accuracy of 92.5\% and an F1-score above 82.5\%, while using far fewer parameters than conventional high-resource deep-learning approaches. The embedded implementation exhibits only marginal performance degradation, with performance drops below 2 percentage points compared to the original full-precision implementation.
These findings demonstrate the feasibility of real-time AI-based heart murmur screening on low-cost microcontrollers. By reducing hardware requirements while preserving predictive performance, the proposed approach can facilitate the development of portable, affordable, and widely deployable diagnostic tools for remote and resource-limited healthcare settings.
\end{abstract}


\section{Introduction}
\label{sec:introduction}

\IEEEPARstart{C}{ardiovascular} diseases (CVDs) confirmed to be the leading cause of death worldwide in the current millennium \cite{whocvds}. The prevalence of CVDs is rising also in low-middle income countries, raising the interest of the scientific community in cost-effective CVD screening methods \cite{sharma2022cost}. Digital auscultation is an interesting candidate, particularly in low-resource scenarios, where access to advanced diagnostic equipment and specialized clinical staff is typically limited. Heart sound recording technology is portable and low-cost: unlike traditional image-based diagnostic technology, such as echocardiography or cardiac computed tomography, electronic stethoscopes can be easily integrated into primary care or even community or domiciliary settings.\\
In particular, the presence of heart murmurs has previously been proven to be an effective biomarker for the detection of Congenital Heart Diseases (CHD), which play an important role in pediatric healthcare \cite{Ainsworth1999, Newburger1983, Ewer2011}. Murmurs are generated by turbulent flow within the heart chambers or large vessels which often arise as a consequence of structural abnormalities such as septal defects, stenosis, or regurgitations \cite{braunwald1992textbook}.\\
The potentiality of heart murmurs analysis for CHD screening in low-resource scenario resulted, in 2022, in the release of the CirCor Digiscope dataset \cite{circor_dataset}. CirCor is currently the largest publicly available dataset of pediatric heart sound recordings, acquired in low-resource settings in rural Brazil \cite{oliveira2021circor}. The dataset includes recordings performed in real-life scenarios: noise, low-quality signals, and label uncertainty make the dataset a benchmark for the development of murmurs detection methods that could be deployed in real low-resource scenarios.\\
Murmur detection on CirCor was the object of the George B. Moody PhysioNet Challenge 2022 \cite{challenge}, stimulating the design of a high number of data-driven approaches for heart murmurs detection and classification. Even if previous works achieved good performance in the detection task, existing literature mostly overlooked a crucial strategical aspect for the deployment of the methods in a real low-resource scenario: the availability of high-resource or cloud computing cannot be taken for granted in these settings due to the lack of internet connection and the unavailability of powerful computers \cite{adedinsewo2025contextual,dangi2025transforming}. 
Therefore, low-resource methods that can be deployed on the edge (on board the recording device itself) should be considered a preferable choice \cite{hartmann2022edge}.\\
In this context, the objective of this work is to develop an AI-based approach for heart murmur detection that can operate in a resource-constrained environment. Unlike most existing studies, which prioritize predictive performance through increasingly complex architectures, the proposed methodology adopts a resource-aware design strategy from the outset.
Specifically, an STM32F401-based development board, characterized by low cost (approximately \$20) and limited computational resources (84~MHz clock frequency, 512~kB Flash memory, and 96~kB RAM), was selected as the target deployment platform before the machine learning model design phase. Consequently, the CNN architecture was developed in Python while explicitly considering the memory and computational constraints imposed by the target MCU.
Furthermore, the complete processing pipeline, including feature extraction, model inference, and prediction aggregation, was deployed on the embedded platform. To enable real-time execution, fixed-point signal processing and model quantization were adopted throughout the implementation. The proposed methodology is thoroughly evaluated in terms of predictive performance, memory footprint, computational complexity, and execution latency, with the ultimate goal of demonstrating the feasibility of real-time heart murmur detection on a low-cost edge device.\\
% In this context, the goal of this work is to propose an AI-based approach for heart murmurs detection in heart sounds in a low-resource scenario with a focus on:
% \begin{itemize}
    % \item Objective 1: the comparison of solutions with different computational cost, with the aim of providing a performative yet cost-effective solution that can be, in the future, implemented on the edge.
    % \item Objective 2: the use of an uncertainty analysis aimed at providing the user with a confidence score regarding the murmur detector to enable selective classification.
% \end{itemize}
%
The rest of the manuscript is organized as follows. Section \ref{sec:background} provides an overview of existing methods, with particular attention to lightweight classifiers and model complexity. Section \ref{sec:mms} presents our proposed approach, including feature extraction, CNN architecture, and MCU deployment; results are showcased in Section \ref{sec:results}. Section \ref{sec:discussion} discusses the predictive performance of both the Python and MCU implementations, together with the current limitations and potential future directions.

\section{Background}
\label{sec:background}

\subsection{Heart murmur detectors}
\label{sec:relatedworks}
A large number of works previously tackled the problem of detecting heart murmurs on the CirCor dataset, stimulated, in the first place, by the George B. Moody PhysioNet Challenge 2022 \cite{challenge}. 
Among the winners, the CUED Acoustics team proposed a hybrid recurrent neural network with Gate Recurrent Unit (GRU) and hidden semi-Markov model (HSMM) algorithm that can both segment the signal and detect a heart murmur \cite{mcdonald2024recurrent}. This first stage receives as input the spectrogram of the recording and produces four segmentations using four parallel HSMMs. Finally, the confidence of the parallel segmentation is compared to infer the most probable segmentation, murmur prediction, and estimate on the signal quality  \cite{mcdonald2024recurrent}. HearTech+ was a co-winner of the Challenge \cite{challenge_cowinner2}. Their system employs a Hierarchical Multi-Scale Convolutional Network (HMS-Net) to perform both murmur detection and clinical-outcome classification. %To control label noise, they apply a lightweight signal-quality screening before training/inference: segments with insufficient low-frequency energy are relabeled as unknown. The rationale is that normal heart sounds concentrate most of their energy in the 20–200 Hz band, while murmurs typically contribute less energy there \cite{Choi2010}. Consequently, they defined a quality ratio as the ratio of spectral power in 20–200 Hz to the full 0–1000 Hz band and used this metric to gate questionable segments. After the quality assessment, a multi-scale spectrogram of the recording (1x, 0.5x, 0.25x) is provided as input to a ResNet. \\  
During the challenge as well as later on, most previous works preferred DL approaches over Machine Learning (ML) \cite{reyna2023heart}, mainly due to the inherent difficulty in handcrafting relevant features from sounds. The use of DL shifted the focus from computationally expensive signal segmentation steps towards the representation of the signal in different domains.\\ 
The proposed input  representations can be classified into two families: A) time-frequency representation, including Short-Time Fourier Transform (STFT) \cite{fakhry2025hybrid,guan2025lachest,orozco2025deep,fernando2024machine,patwa2025heart,mcdonald2024recurrent,han2024deep} and Continuous Wavelet Transform (CWT) \cite{vimalajeewa2025multiscale,fakhry2025hybrid,funnelnet}; and B) spectral features, such as Mel Frequency Cepstral Coefficients (MFCCs) \cite{safdar2025empowering,orozco2025deep,kalimuthu2025comparative,patwa2025heart,ozcan2024rapid} and Mel Spectrogram \cite{han2024deep,luo2024phonocardiogram}. A few previous works achieved good performances by applying DL directly on the signal (or sub-segments) in the time domain \cite{morshed2025deep,shin2025temporal}.\\
Concerning the selection of the DL model, the majority of the previous works aiming at keeping the computational cost to reasonable levels relied on Convolutional Neural Networks (CNN), either pure \cite{safdar2025empowering,han2024deep,morshed2025deep, funnelnet, tinyml_systolic} or combined with recurrent layers \cite{guan2025lachest}, attention layers \cite{kalimuthu2025comparative} or LSTM \cite{fakhry2025hybrid}. Other successful approaches include Recurrent Neural Networks (RNN) \cite{mcdonald2024recurrent} and transformers \cite{niizumi2024exploring}.

\subsection{Challenges and motivation for on the edge computing}

Most classifiers today rely on DL models, whose complexity may vary quite significantly depending on the architecture: implementation on the edge may be challenging if not designed from the beginning. In this scenario, a few state of the art works focused on the deployability of their proposed heart murmurs detectors on microcontroller-based embedded systems. \\
Jobayer et al.~\cite{funnelnet} proposed an end-to-end lightweight architecture based on CWT features and Depthwise Separable Convolutions (DSC), requiring only approximately 5.4k parameters. The authors deployed a real-time TFLite implementation on Raspberry Pi and Android devices, reporting average accuracies of 91\% and 80\%, respectively. However, it is not explicitly stated whether the feature extraction stage is also executed on the embedded platforms, and the on-device profiling mainly focuses on model inference.
Similarly, Alanes et al.~\cite{tinyml_systolic} presented a TinyML framework based on Mel Frequency Energy features and compact CNNs, with a model requiring approximately 43k parameters. Their model was deployed through the Edge Impulse platform, achieving test accuracies above 92\%, RAM usage below 90~kB, and an inference latency of approximately 1.1~s on compatible embedded platforms, such as ESP32- and Arduino-based devices.
Safdar et al.~\cite{safdar2025empowering} proposed SET-MobileNet, a lightweight architecture for heart sound classification based on MobileNetV2. Their approach employs multimodal acoustic features, including log-mel spectrograms, MFCCs and chroma features. They report fast inference on mobile and embedded platforms, including Raspberry Pi. Nevertheless, this model still requires substantially higher resources than low-cost MCUs, with approximately 2.25M parameters and an 8.61~MB model size.\\
Despite these advances, existing works mainly focus on model-level optimization and inference deployment, while the complete embedded processing chain is rarely evaluated under realistic MCU constraints. In this study, we provide a comprehensive analysis of the full pipeline, including fixed-point feature extraction accuracy, memory footprint, execution latency and quantized model inference.\\

\section{Materials and Methods}
\label{sec:mms}

\subsection{Description of the dataset}
The CirCor DigiScope Phonocardiogram Dataset \cite{circor_dataset}, contains a large number of pediatric heart-sound recordings collected during two mass-screening campaigns in Brazil in 2014 and 2015. % It includes subjects younger than 21 years who presented for screening and completed a standardized clinical workflow (history/physical, nursing assessment, and when indicated cardiac investigations). 
The dataset contains 5272 phonocardiogram (PCG) recordings from 1568 subjects (mean age $6.1 \pm 4.3$ years; range $0-21$ years). All PCGs were recorded using the Littmann 3200 electronic stethoscope and with a sample rate of 4000 Hz. Individual recordings last $4.8–80.4$ s (mean $22.9 \pm 7.4$ s), with more than 33.5 h of audio. Multiple recordings per subject are available from the standard auscultation sites (mitral, tricuspid, aortic and  pulmonic). The dataset contains a subject description file and segmentation file for every recording, containing segmentation information regarding the start and end points of the fundamental heart sounds S1 and S2. The CirCor dataset %is the largest available pediatric dataset, 
aimed to support research on murmur detection. Indeed, every audio recording is annotated with three possible labels: murmur \textit{Absent}, murmur \textit{Present} or \textit{Unknown}. Recordings marked as \textit{Unknown} by the annotator represent signals that did not meet the required signal quality standards, i.e. recordings did not lead to a reliable murmur characterization and description.
\begin{table}[!ht]
    \caption{Distribution of murmur labels across dataset splits. Entries are recordings with unique patients in parentheses.}
    \centering
    \label{tab:class-split}
    \begin{tabular}{cccc}
    \hline
         & Absent & Present & Unknown\\
    \hline
      Train   & 2508 (695) & 499 (179) & 156 (68)\\
      Validation  & 380 (106) & 69 (27) & 37 (16) \\
      Test   & 1249 (343) & 289 (99) & 85 (35) \\
      \hline
      All & 4137 (1144) & 857 (305) & 278 (119) \\
      \hline
    \end{tabular}
\end{table}
The entire dataset is partitioned into training (60\%), validation (30\%) and test sets (10\%). Table \ref{tab:class-split} summarizes the label distribution for the three murmur classes (\textit{Absent}, \textit{Present}, \textit{Unknown}) across the dataset. Counts refer to recordings, with the corresponding number of unique patients shown in parentheses. The dataset is markedly imbalanced: %In total, Absent accounts for 4065/5200 recordings (78.2\%), Present for 857/5200 (16.5\%), and Unknown for 278/5200 (5.3\%). Patient counts sum to 1568 (1144 Absent, 305 Present, 119 Unknown), confirming that multiple recordings per patient are available. 
the \textit{Absent} class represents 78.2\% of all of the recordings, whereas the \textit{Present} and \textit{Unknown} class represent 16.5\% and 5.3\%, respectively.
% This imbalance motivated us to approach the problem in a different way from previous studies, reformulating the classification problem to improve the model performance by maintaining a low level of parameters.

\subsection{Feature extraction}
Frequency-domain features were extracted from the raw audio signals using the STFT. This technique computes the Fast Fourier Transform (FFT) over short, overlapping segments of the signal, after multiplying each segment by a windowing function to reduce spectral leakage. In this work, a Hann window was employed, which is commonly used for this purpose. The segment length, denoted as 
$N_{FFT}$, is typically chosen as a power of two to improve computational efficiency.
The STFT produces a two-dimensional matrix, where the first dimension represents the number of positive frequency bins 
($N_{FFT}/2+1$) and the second dimension corresponds to the time windows. To focus on the clinically relevant frequency range, only the lower half of the frequency bins were retained, effectively applying a low-pass filter with a cutoff at 1 kHz, as higher frequencies generally do not contain meaningful heart sound information \cite{challenge_cowinner2}.
Rather than applying STFT to the entire audio recording at once, the signal was segmented into 2-second windows with a 1-second overlap. This choice ensures that each segment captures at least one complete systolic-diastolic cardiac cycle. As a result, for each patient and recording location, the feature extraction process generates a series of two-dimensional spectrograms, the number of which depends on the duration of the original recording.

\subsection{Model architecture \& training details}
\begin{figure}[!ht]
    \centering
\includegraphics[width=\linewidth]{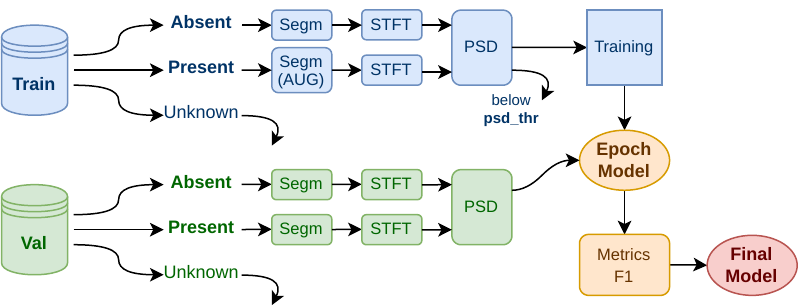}
    \caption{
    Representation of the feature extraction and training pipeline. Raw audio is segmented (samples labeled \textit{Present} are augmented only in the training set), and STFT are computed. Finally, segments with poor PSD are discarded.
    % for each segment. The model with the highest validation F1-score across epochs is selected as final.
    }
    \label{fig:training_schematic}
\end{figure}

\begin{figure}[!ht]
    \centering
\includegraphics[width=0.95\linewidth]{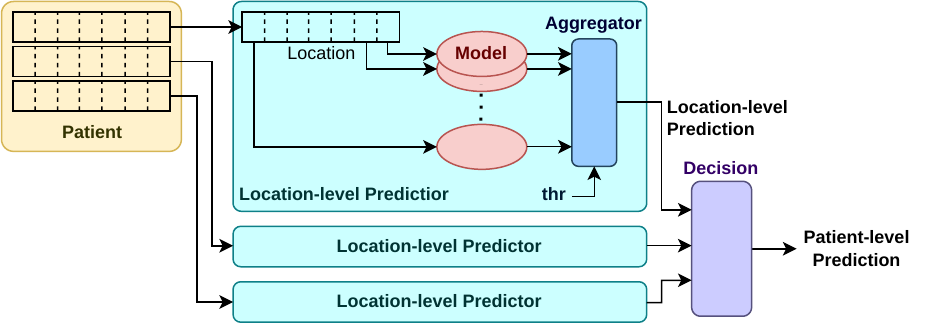}
     \caption{Description of the patient-level prediction. An outcome is produced for each location; if at least one is positive (murmur detected) the patient is inferred to have a murmur.
     }
     
    \label{fig:prediction_schematic}
\end{figure}

A CNN model was developed to detect the presence of cardiac murmurs using the previously extracted 2D  features.
Due to significant class imbalance and variations in audio quality, some design decisions were implemented to improve model performance, as outlined in Fig.~\ref{fig:training_schematic}. 
First of all, patients labeled as \textit{Unknown} were excluded from the training set. This decision, which has also been adopted in previous classification studies \cite{fakhry2025hybrid, guan2025lachest, orozco2025deep, vimalajeewa2025multiscale, ozcan2024rapid, shin2025temporal}, was motivated by the limited number of such cases and the difficulty in distinguishing them from \textit{Known} samples (i.e., those labeled \textit{Absent} or \textit{Present}). The \textit{Unknown} label typically arises from poor audio quality or the presence of confounding conditions that prevent expert annotation. Since the primary goal of this work is to accurately detect murmur presence, the problem was reformulated as a binary classification task (\textit{Absent} vs. \textit{Present}). 
To address the class imbalance, the number of samples belonging to the \textit{Present} class was increased through data augmentation. Specifically, Gaussian noise addition, time stretching, time shifting, gain variation, and the application of random band-pass and band-stop filters were employed to augment the \textit{Present} recordings by a factor of five. 
% The parameters of these transformations were carefully selected to increase signal variability while preserving the acoustic characteristics associated with heart murmurs.
The frequency bands used for the random filters were selected to emphasize low-frequency components (100-800~Hz), where heart sounds and murmurs are more commonly represented. Conversely, the stop-band filters were applied to higher-frequency ranges (500-1000~Hz), which are generally less informative for murmur detection. This strategy increases signal variability while preserving the acoustic characteristics associated with heart murmurs.\\
Following feature extraction, each segment underwent a quality assessment based on its power spectral density (PSD).
The PSD is computed for each segment over two frequency bands: 20–200 Hz and 0–1000 Hz, as shown in Eq. \ref{eq:psd_band}. Here $X$ denotes the STFT matrix, while $b_x$, $b_y$ and $b_n$ represent the frequency bins corresponding to 20 Hz, 200 Hz, and 1000 Hz, respectively. $T$ denotes the last short-time segment.
\begin{equation} \label{eq:psd_band}
PSD_{\text{band}}=\sum_{b=b_x}^{b_y}\sum_{t=0}^{T}\bigl|X_{b,t}\bigr|^2,
\;\;\:
PSD_{\text{tot}}=\sum_{b=0}^{b_n}\sum_{t=0}^{T}\bigl|X_{b,t}\bigr|^2.
\end{equation}
Finally, the ratio is calculated as $PSD_{\text{ratio}}=P_{\text{band}}/P_{\text{tot}}$.
%Eq. \ref{eq:psd_ratio}.
%\begin{equation}\label{eq:psd_ratio}
%\PSD_{\text{ratio}}=\frac{P_{\text{band}}%}{P_{\text{tot}}}.
%\end{equation}
The 20–200 Hz frequency range was selected since the systolic and diastolic components of heart sounds exhibit the highest energy within this band.
Segments with a $PSD_{\text{ratio}}$ below a predefined threshold \textit{psd\_thr} were considered low-quality and discarded. To ensure that data from each recording location remained available, at least the five highest-quality segments per location were retained, even if their $PSD_{\text{ratio}}$ values were below the threshold.\\
% \begin{table}[!ht]
%     \centering
%     \caption{The Architecture of the Proposed Model}
%     \begin{tabular}{c c c c c} 
%          \multicolumn{5}{c}{\textbf{LIGHT}}\\
%          \hline
%          \textbf{Layer} & \textbf{In CH} & \textbf{Out CH} & \textbf{Activation} & \textbf{Dropout}\\
%          \hline
%          Conv2d+MaxPool2d       & 1     & 16    & ReLU      & 0.1\\
%          Conv2d+MaxPool2d       & 16    & 32    & ReLU      & 0.1\\
%          Conv2d+AdAvgPool2d       & 32    & 64    & ReLU      & 0.1\\
%          Linear             & 64    & 2     & Softmax   & - \\
%          \hline
%          \multicolumn{5}{l}{For all \textit{Conv2d} layers, a 3x3 kernel with a padding of 1 is used, while}\\
%          \multicolumn{5}{l}{the \textit{MaxPool2d} layers employed 2x2 kernels. The \textit{AdAvgPool2d} layers } \\
%          \multicolumn{5}{l}{correspond to AdaptiveAvgPool2d with an output size of 1.}\\
%          \multicolumn{5}{l}{Columns \textit{In CH} and \textit{Out CH} represent input and output channels.}\\
%     \end{tabular}
%     \label{tab:model_arch}
% \end{table}
\begin{table}[h]
    \centering
    \caption{The Architecture of the Proposed Model}
    \begin{tabular}{c c c c c c} 
         \hline
         \textbf{Layer} & \textbf{Input} & \textbf{Output} & \textbf{Kernel} & \textbf{Activation} & \textbf{Dropout}\\
         \hline
         Conv2d             & 1     & 16    & 3 & ReLU      & 0.1\\
         MaxPool2d          & 16    & 16    & 2 & -         & - \\
         Conv2d             & 16    & 32    & 3 & ReLU      & 0.1\\
         MaxPool2d          & 32    & 32    & 2 & -         & - \\
         Conv2d             & 32    & 64    & 3 & ReLU      & 0.1\\
         AvgPool2d        & 64    & 64    & 1 & -         & - \\
         Linear             & 64    & 2     & - & Softmax   & - \\
         \hline
    \end{tabular}
    \label{tab:model_arch}
\end{table}
The selected model is a lightweight CNN composed of three convolutional layers, followed by a fully connected classification layer and a softmax output. It was first implemented in Python using the PyTorch framework with the scope of validating the model and tuning the hyperparameters. Details of the model architecture are reported in Table~\ref{tab:model_arch}, while the training hyperparameters are summarized as follows: the loss function used is \textit{CrossEntropyLoss}, optimization is performed with the \textit{AdamW} optimizer, the learning rate is set to $10^{-3}$ and the model is trained for 50 epochs.\\
After training, the model epoch with the highest F1-score on the validation set was selected as the final model. The validation set underwent the same pre-processing steps as the training data, except that no data augmentation or low-PSD segment rejection was applied.\\
The trained CNN outputs a binary prediction for individual 2-second segments. Patient-level classification is then derived by aggregating predictions across all segments and recording locations. Fig.~\ref{fig:prediction_schematic} displays the process of the final prediction. 
Each segment is first classified individually. Then, predictions from the same recording location are combined by an \textit{Aggregator} block: the softmax probabilities associated with the \textit{Present} class are averaged across all segments, and a threshold value, denoted as \textit{thr}, is applied to classify the entire recording location.
% if the ratio of \textit{Present} over \textit{Absent} exceeds the value "thr", the location is classified as \textit{Present} by the "Location Predictor" module. 
This module, called "Location-level Predictor", is repeated for all recording locations.
Finally, a "Decision" block analyzes the location-level results: if at least one location is classified as \textit{Present}, the patient is predicted to have a murmur.\\
The threshold \textit{thr} is calibrated after the training phase using the validation set, with the goal of maximizing the F1-score at the location level.
% In the end, the same threshold value was selected for all three models, with the final value set to 40\%. This value was chosen as a trade-off among the three architectures, representing an intermediate setting that balances their individual performance trends.
Accordingly, the test set remains completely unseen during all stages of training and post-processing.

\subsection{Real-Time Feature Extraction and On-The-Edge Implementation}
A real-time implementation of the proposed pipeline was developed on an STM32F401-NUCLEO board.\\
Audio samples were streamed through the UART peripheral and acquired using Direct Memory Access (DMA) to minimize CPU overhead. Feature extraction and signal processing operations were implemented using the CMSIS-DSP library \cite{cmsis-dsp}, while the CNN model was converted into optimized C code using STM32Cube.AI \cite{st_cube_ai}.
Prior to deployment, Post-Training Quantization (PTQ) was applied to convert the network weights from floating-point representation to the \textit{int8} data type, reducing both memory requirements and inference latency. A subset of the training dataset was used as a calibration set to determine the scaling factors required for input feature quantization.\\
Some design choices were adopted to optimize real-time execution on the MCU, ensuring that feature extraction and inference could be completed before the acquisition of the next audio segment. The segment duration was fixed to 2.048~s, corresponding to 8192 audio samples. With an overlap of 4096 samples, the second half of each segment coincides exactly with the first half of the subsequent one. Consequently, only the STFT coefficients associated with the newly acquired samples must be computed, reducing the feature extraction workload by approximately 50\%.
\begin{figure}
    \centering
    \includegraphics[width=\linewidth]{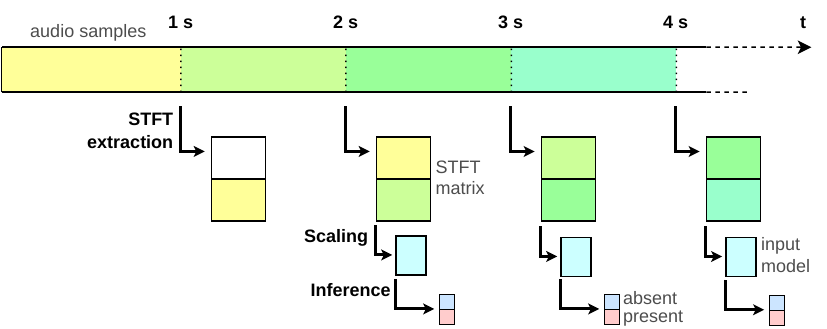}
    \caption{Real-time MCU processing pipeline. The feature matrix is updated by retaining the overlapping STFT coefficients and computing the new coefficients. The STFT matrix is scaled and provided as input to the CNN model to compute murmur probability.}
   \label{fig:real-time}
\end{figure}
Fig.~\ref{fig:real-time} illustrates the processing flow executed during audio streaming. After the acquisition of 4096 new samples, the STFT matrix is updated by shifting the coefficients associated with the overlapping portion of the previous segment and computing the coefficients corresponding to the newly received samples.
Before inference, the feature matrix is scaled and quantized to the \textit{int8} format expected by the deployed model. The softmax probability associated with the \textit{Present} class is accumulated across all segments belonging to the same recording location. 
Once all segments from a location have been processed, a location-level prediction is obtained by applying the threshold value \textit{thr} to the aggregated score.\\
The same procedure is repeated for all recording locations, and the resulting location-level predictions are finally combined to produce the patient-level diagnosis.
In summary, the following pipeline was implemented on the MCU:
\begin{enumerate}
    \item \textbf{Data streaming}: audio samples are received by the device in bursts of 4096 samples, corresponding to approximately 1~s;
    \item \textbf{Feature extraction}: STFT features are computed in \textbf{Q15}, and are combined to form the feature matrix;
    \item \textbf{Model inference}: the feature matrix is scaled to \textbf{int8}, the quantized model is executed, and the two output classes are dequantized. Finally, the softmax probabilities are computed;
    \item \textbf{Murmur prediction}: once all samples from a recording location have been processed, the murmur-level prediction is computed;
    \item \textbf{Patient prediction}: once all recording locations have been processed, the final patient-level prediction is computed.
\end{enumerate}

\subsection{Evaluation metrics}

Model performance was assessed using standard metrics for binary classification, considering the two classes: murmur \textit{Present} and murmur \textit{Absent}. Specifically, we report the classification accuracy, which measures the overall proportion of correctly classified samples, and the F1-score, defined as the harmonic mean of precision and recall, providing a balanced measure of performance in the presence of class imbalance. Precision quantifies the proportion of correctly identified positive cases among all samples predicted as positive, while the Recall (or Sensitivity) measures the proportion of correctly identified positive cases among all actual positives. These metrics were computed at the location level and then aggregated to assess overall model performance on a patient-level basis.

\section{Results}
\label{sec:results}

\subsection{MCU feature extraction: accuracy and profiling}

Computing spectrograms, and consequently STFT coefficients, on an MCU requires both computational resources and memory for intermediate buffers and lookup tables. Due to the limited hardware resources and real-time constraints of the target platform, only fixed-point implementations were considered. In particular, two numerical formats were evaluated: a fixed point format with 15 fractional units, \textbf{Q15}, and with 31 fractional units, \textbf{Q31}.\\
The accuracy of the embedded implementation was assessed by comparing the extracted STFT coefficients with those obtained from the floating-point Python implementation. The floating-point coefficients was used as the reference, and the absolute error was computed for both \textbf{Q15} and \textbf{Q31} formats over the entire dataset, including training, validation, and test sets. 
Since \textbf{Q31} values provide 16 additional fractional bits compared to \textbf{Q15}, the errors were right-shifted by 16 bits to enable a direct comparison in \textbf{Q15} LSB units.
From the resulting error distribution, the following metrics were computed:
\begin{itemize}
    \item Mean Absolute Error (MAE)
    \item Root Mean Square Error (RMSE)
    \item Percentile 95 (P95)
    \item Maximum error (Max)
\end{itemize}
% Since \textbf{Q31} values provide 16 additional fractional bits compared to \textbf{Q15}, the absolute error was right-shifted by 16 bits to enable a direct and fair comparison in LSB units.\\
\begin{table}[!ht]
    \centering
    \caption{Error metrics of the MCU fixed-point STFT implementation}
\begin{tabular}{c c c c c c}
     \hline
     & \textbf{MAE} & \textbf{RMSE} & \textbf{P95} & \textbf{Max}\\
    \hline
     \textbf{Q15} & 0.5068 & 0.5889 & 0.9617 & 4.1474 \\
     \textbf{Q31} & 9.521e-4 & 7.663e-3 & 3.011e-3 & 0.9963 \\
    \hline
\end{tabular}
\label{tab:feature_cmsis}
\end{table}
As expected, the \textbf{Q31} format provides higher numerical precision, with error metrics that are approximately three orders of magnitude lower than those obtained with \textbf{Q15}. However, this improvement comes at the cost of doubling the memory required to store the feature buffers.
For the \textbf{Q15} representation, the MAE is close to the theoretical quantization error (0.5 LSB), indicating that most of the observed error originates from the fixed-point representation rather than from inaccuracies in the algorithm itself. The P95 metric shows that the typical worst-case error is approximately 1 LSB. Although the maximum observed error reaches 4 LSBs, such events occur only in a negligible fraction of the samples.
Similarly, the RMSE remains below 0.6 LSB, confirming that the overall numerical accuracy of the \textbf{Q15} implementation is largely limited by quantization effects rather than implementation errors.\\
\begin{table}[!ht]
    \centering
    \caption{Profiling results of the MCU feature extraction implementation}
\begin{tabular}{c c c c c}
     \hline
     & \multirow{2}{*}{\textbf{Ram [kB]}} & \multirow{2}{*}{\textbf{Flash [kB]}} & \multicolumn{2}{c}{\textbf{Time$^\alpha$ [ms]}}\\
     & & & avg & std\\
    \hline
     \textbf{Q15} & 16.25 & 51.13 & 95.30 & 0.10 \\
     \textbf{Q31} & 32.5 & 82.5 & 114.787 & 0.036 \\
    \hline
    \multicolumn{5}{l}{$^\alpha$Time is averaged along the audio of test dataset}\\
\end{tabular}
\label{tab:feature_profiling}
\end{table}
Table~\ref{tab:feature_profiling} reports the execution time and memory footprint measured on the target MCU. While the additional processing time required by the \textbf{Q31} implementation is relatively small (approximately 20 ms) compared to the acquisition time of half a segment (about 1 s), the increased memory consumption represents a more significant limitation.
In particular, RAM usage becomes critical when considering that a substantial portion of memory must also be allocated to the CNN activations during inference.\\
Considering both numerical accuracy and resource utilization, the \textbf{Q15} format was selected for the final implementation. Although \textbf{Q31} provides higher precision, the results demonstrate that \textbf{Q15} maintains sufficient numerical accuracy for the target application while significantly reducing memory requirements.

\subsection{Model selection and performance}
\begin{figure}[!ht]
  \centering
  \subfloat[]{%
\includegraphics[width=.5\linewidth]{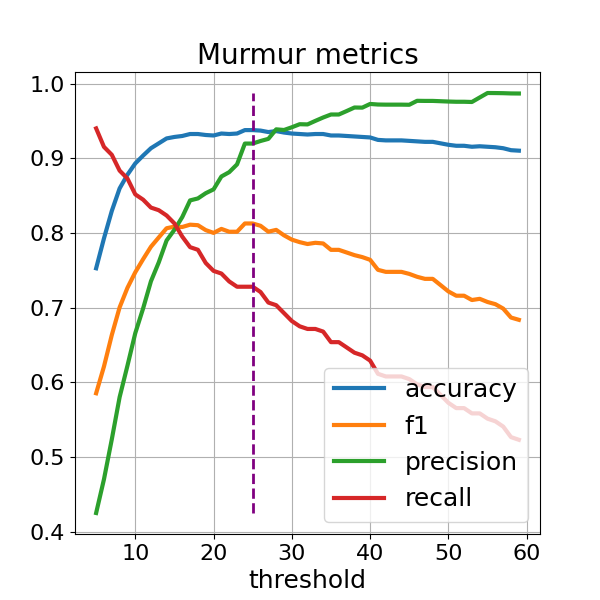}}
  \hfill
  \subfloat[]{%
\includegraphics[width=.5\linewidth]{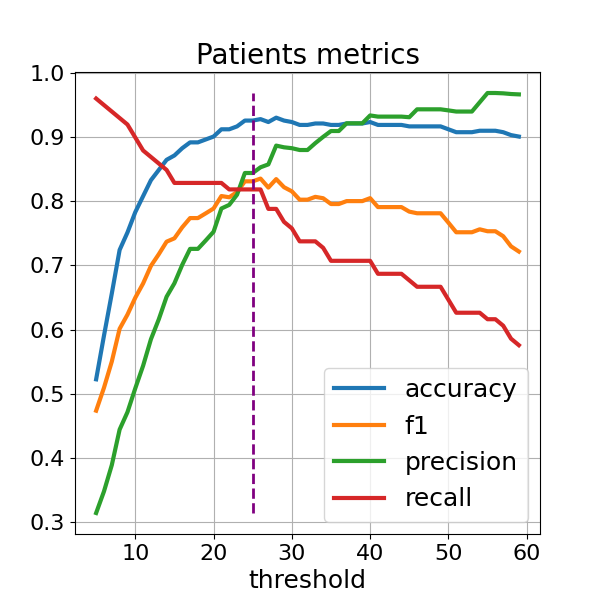}}
\caption{Metrics of Python predictions for test dataset by varying the threshold for murmur-level (a) and patient-level (b). The vertical dashed line indicates the calibrated threshold from validation dataset.}
\label{fig:metrics_thr_py}
\end{figure}

The impact of the $N_{FFT}$ and $psd\_thr$ parameters was evaluated using a 5-fold cross-validation (CV) procedure. To ensure a realistic validation scenario, the folds were created at the patient level. Specifically, the 874 labeled \textit{Absent} or \textit{Present} patients in the training set were randomly divided into five folds, and in each iteration, the model was trained on four folds and evaluated on the remaining one. This patient-level split prevents data leakage by ensuring that segments from the same patient do not appear in both the training and validation sets, which could otherwise bias the CV results.
\begin{figure}[h]
  \centering
  \includegraphics[width=\linewidth]{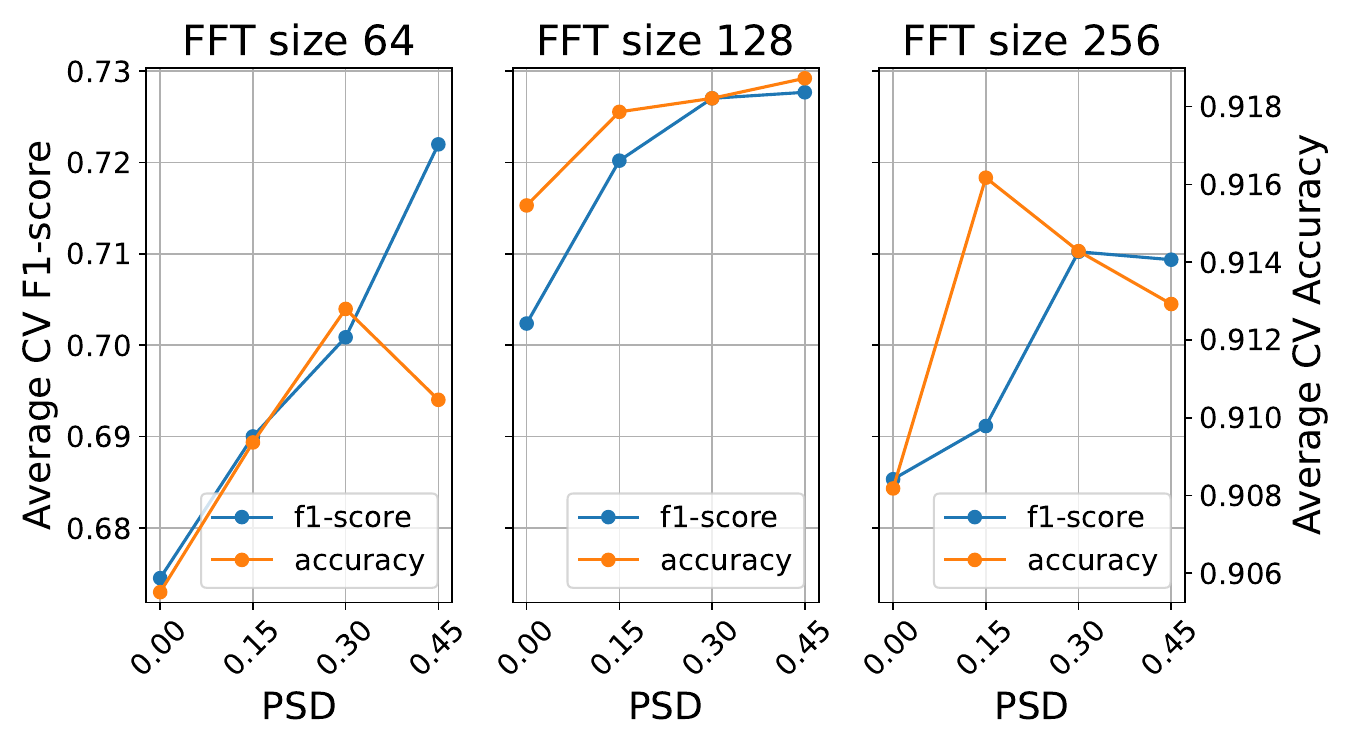}
  \caption{Cross-validation results for the model with three different $N_{fft}$ values: 64, 128, and 256. Each graph shows the average Accuracy and F1-score across the 5 folds for different \textit{psd\_thr} values.}
  \label{fig:cv_result}
\end{figure}\\
Three values of $N_{FFT}$ (64, 128, and 256) were considered. These values were selected because, given the relatively low sampling rate, increasing the FFT window beyond 256 does not improve spectrogram quality, while reducing it below 64 significantly degrades spectral resolution. Regarding the second parameter, the maximum PSD threshold tested was 0.45. Higher thresholds were not considered, as they would result in excessive segment rejection, potentially compromising the subsequent aggregation step.
The CV results are reported in Fig.~\ref{fig:cv_result}, which shows the average F1-score and accuracy across all folds. It is important to note that these metrics refer exclusively to segment-level classification, without including the final patient-level aggregation.\\
Based on the CV results, the parameters $N_{FFT}=128$ and $psd\_thr=0.45$ were selected for the final model, as they obtain the best F1-score.
A new CNN model was then trained on the full training dataset following the complete pipeline to generate patient-level predictions.
The overall performance metrics are reported in Table~\ref{tab:model_final_result} in the \textit{Python} row. 
\textit{Murmur} metrics refer to the prediction obtained for a single recording location, whereas the \textit{Patient} metrics represent the final patient-level prediction obtained after aggregation.
Fig.~\ref{fig:metrics_thr_py} shows the variation of the performance metrics as a function of the aggregation threshold for murmur-level prediction (Fig.~\ref{fig:metrics_thr_py}(a)) and patient-level prediction (Fig.~\ref{fig:metrics_thr_py}(b)). The final threshold was calibrated on the validation dataset by maximizing the F1-score and was set to 25\%.

\begin{table}[!h]
\caption{Metrics on the test set for the final model}
    \label{tab:model_final_result}
    \centering
    \begin{tabular}{c c c c c c c}
    \hline
    &  & \textbf{Accuracy} & \textbf{F1-score} & \textbf{Precision} & \textbf{Recall}\\
    \hline
    \multirow{2}{*}{Python} & Murmur  &  93.77\% & 81.26\% & 91.96\% & 72.79\%\\
    & Patient &  92.53\% & 83.08\% & 84.38\% & 81.82\%\\
    \hline
    \multirow{2}{*}{PTQ} & Murmur  &  93.64\% & 80.56\% & 93.06\% & 71.02\%\\
    & Patient &  93.21\% & 84.38\% & 87.10\% & 81.82\%\\
    \hline
    \multirow{2}{*}{MCU} & Murmur  &  93.50\% & 80.00\% & 93.40\% & 69.96\%\\
    & Patient &  92.53\% & 82.54\% & 86.67\% & 78.79\%\\
    \hline
    \end{tabular}
\end{table}

\subsection{MCU model performance and profiling}

\begin{figure}[!ht]
  \centering
  \subfloat[]{%
\includegraphics[width=.5\linewidth]{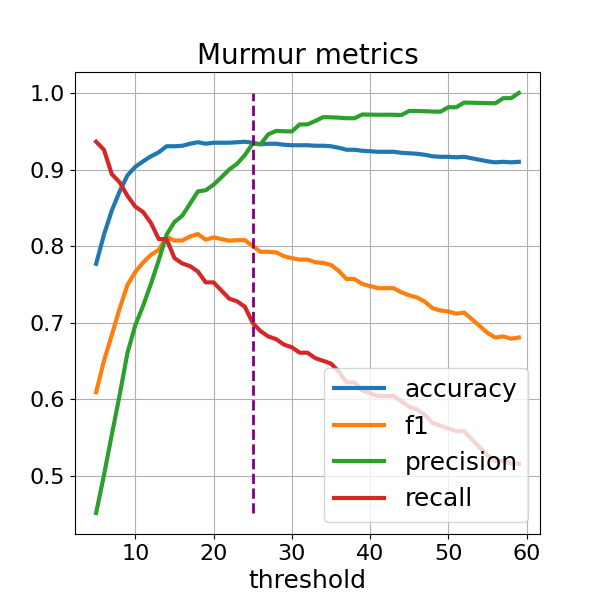}}
  \hfill
  \subfloat[]{%
\includegraphics[width=.5\linewidth]{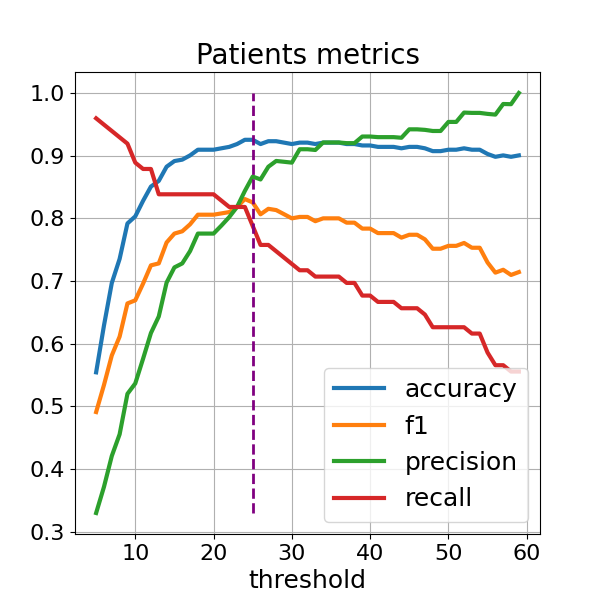}}
\caption{Metrics of MCU predictions for test dataset by varying the threshold for murmur-level (a) and patient-level (b). The vertical dashed line indicates the calibrated threshold.}
\label{fig:metrics_thr_c}
\end{figure}

The \textit{ST Edge AI Developer Cloud} \cite{st_ai} online tool, developed by STMicroelectronics, was used to analyze and quantize the proposed CNN model.
The required hardware resources are reported in Table~\ref{tab:on-the-edge}. 
\begin{table}[!h]
    \caption{MCU model analysis report}
    \label{tab:on-the-edge}
    \centering
    \begin{tabular}{c c c c c}
    \hline
    & \textbf{Params [kB]} & \textbf{Flash [kB]} & \textbf{Ram [kB]} & \textbf{MACC}\\
    \hline
    Native & 91.51 & 107.72 & 101.25 & 10.03 M\\
    Quantized & 23.21 & 49.75 & 32.25 & 9.91 M\\
    \hline
    \end{tabular}
\end{table}
The \textit{Flash} field includes the model weights (Params) and the compiled library code, while the computational complexity is expressed in terms of multiply and accumulate operations (MACCs). Both the native and quantized model configurations are reported. The results show that the native model cannot be deployed on the target MCU due to insufficient RAM.
The metrics obtained with the quantized model are very close to those of the native model, with only a negligible drop (\textit{PTQ} row of Table~\ref{tab:model_final_result}.)\\
% The following pipeline was implemented on the MCU:
% \begin{enumerate}
%     \item \textbf{Data streaming}: audio samples are received by the device in bursts of 4096 samples, corresponding to approximately 1~s;
%     \item \textbf{Feature extraction}: STFT features are computed in \textbf{Q15}, and are combined to form the feature matrix;
%     \item \textbf{Model inference}: the feature matrix is scaled to \textbf{int8}, the quantized model is executed, and the two output classes are dequantized. Finally, the softmax probabilities are computed;
%     \item \textbf{Murmur prediction}: once all samples from a recording location have been processed, the murmur-level prediction is computed;
%     \item \textbf{Patient prediction}: once all recording locations have been processed, the final patient-level prediction is computed.
% \end{enumerate}
The murmur and patient prediction performances obtained by the process described in Sec.~\ref{sec:mms}.D are reported in the \textit{MCU} row of Table~\ref{tab:model_final_result}, while the metrics in function of the threshold are shown in Fig.~\ref{fig:metrics_thr_c}.
To ensure real-time execution, steps 2 and 3 must be completed within the 1-s interval required to acquire the next burst of samples.
The feature extraction time is reported in Table~\ref{tab:feature_profiling}, while the model inference time was measured over all segments of the test dataset. The average inference time was 245.242~ms, with a standard deviation of 0.022~ms.
Combining the average feature extraction and inference times results in a total processing time of 340.54~ms. Therefore, the proposed implementation provides a time margin of approximately 683~ms before the next audio burst is available, confirming its suitability for real-time execution on the target MCU.

\section{Discussion \& Conclusions}
\label{sec:discussion}

\subsection{Feasibility of Real-Time murmur detection on Low-Cost Edge Devices}

The main effort of this work is to propose a method that achieves an optimal balance between detection accuracy, sensitivity and computational cost. Developing a lightweight solution is essential to enable deployment on the edge, which is a known limitation of many AI-based systems. \\
To address these constraints, a compact CNN architecture was designed and trained while explicitly considering the hardware limitations of the target MCU. The proposed aggregation strategy combines segment-level predictions to obtain location-level and patient-level diagnoses. 
The proposed approach achieves a murmur-level accuracy of 93.5\%. However, due to the class imbalance, the F1-score for the \textit{Present} class was considered a more informative performance metric. Accordingly, the high Precision of 93.4\% and F1-score of 80.0\% indicate reliable murmur detection.
% The proposed approach achieves a murmur-level F1-score of 80\%, together with high precision of 93.4\%. 
Moreover, patient-level aggregation further improves both the F1-score and Recall, which are particularly relevant metrics in screening applications, where missed detections may have significant clinical consequences.
Fig.~\ref{fig:metrics_thr_py} and Fig.~\ref{fig:metrics_thr_c} highlight the importance of the aggregation threshold. Proper calibration of this parameter allows an effective balance between Precision and Recall, resulting in the highest F1-score.
The threshold calibrated on the validation dataset achieves an F1-score close to the optimum observed for patient-level prediction, as shown in Fig.~\ref{fig:metrics_thr_py}(b) and Fig.~\ref{fig:metrics_thr_c}(b). A similar behavior is observed for murmur prediction. Although in Fig.~\ref{fig:metrics_thr_c}(a) the selected threshold is slightly farther from the optimal value, the resulting F1-score reduction remains below 1\%. These results suggest that both the CNN model and the aggregation procedure generalize well to unseen patients. 
A second major contribution of this work is the demonstration that the complete processing chain can be executed on a resource-constrained MCU. The results show that STFT feature extraction can be efficiently implemented using fixed-point arithmetic while maintaining sufficient numerical accuracy. In particular, the \textbf{Q15} representation was found to provide an effective compromise between precision and memory consumption, enabling substantial RAM savings with negligible impact on feature quality.
The performance degradation introduced by the embedded implementation remains limited. At the patient level, the F1-score decreases by only 0.5\% compared to the Python implementation, while the murmur-level F1-score exhibits a reduction below 1.3\%. Although Recall is slightly reduced, Precision remains comparable or even slightly improved, confirming that the fixed-point implementation and model quantization preserve most of the predictive capability of the original model. \\
Table~\ref{tab:stateoftheart} summarizes representative state-of-the-art models, reporting network complexity when available, classification accuracy, and whether an embedded implementation was evaluated.
Compared with state-of-the-art approaches, the proposed solution achieves competitive classification performance while reducing the model size by several orders of magnitude. Combined with the optimized fixed-point feature extraction strategy and the lightweight CNN architecture, this reduction enables real-time execution on a low-cost MCU platform without requiring external computing resources.
\begin{table}[h]
\caption{Comparison against existing heart murmurs classifiers using CirCor.}
    \label{tab:stateoftheart}
    \centering
    \begin{tabular}{c c c c c}
    \hline
    \textbf{Reference} & \textbf{Classifier} & \textbf{On-Edge} & \textbf{Accuracy} & \textbf{Params}\\
    \hline
    Niizumi \cite{niizumi2024exploring} & Transformer & No & 83.2\% & 85.4M \\
    Han \cite{han2024deep} &  CNN & No & 85.9\% & 141.5k \\
    McDonald \cite{mcdonald2024recurrent} & RNN+HSMM & No & 77.6\% & N/A \\
    Kalimuthu \cite{kalimuthu2025comparative} & CNN+attention & No & 95.2\% & N/A \\
    Guan \cite{guan2025lachest}  & CRNN & No & 92.4\% & 43.6M \\
    Safdar \cite{safdar2025empowering}  & CNN & Complete & 99.9\% & 2.25M \\
    Fakhry \cite{fakhry2025hybrid}    & CNN+LSTM & No & 90.2\% & 3.4M \\
    Morshed \cite{morshed2025deep}    & CNN & No & 92.2\% & N/A \\
    Jobayer ~\cite{funnelnet} & DSC & Model & 91.0\% & 5.4k \\
    Alanes ~\cite{tinyml_systolic} & CNN & Complete & 94.0\% & 43k \\
    Ours & CNN & Complete & 92.5\% & 23k \\
    \hline
    \end{tabular}
\end{table}
Overall, the results demonstrate that accurate murmur detection can be achieved on affordable embedded hardware, e.g., directly on a digital stethoscope, supporting the feasibility of portable and low-cost screening devices for point-of-care and remote healthcare applications.

\subsection{Limitations and Future Directions}
Despite the promising results, several limitations should be acknowledged. First, the study was conducted exclusively on the CirCor dataset considering the binary murmur detection. The dataset, although large and diverse, represents a specific pediatric population and acquisition setting. This limits the generalizability of the results to other age groups, recording devices, and environmental conditions. 
Moreover, the present analysis does not include energy or power consumption measurements. These evaluations are essential for the development of battery-powered devices, which would further improve the portability and practical usability of the final application.
% Moreover, the dataset annotations inherently contain label noise, particularly for the \textit{Unknown} class, which may affect both training and evaluation reliability.
\\
Future research will focus on extending the present framework along several directions. First, model generalization will be evaluated through external validation on independent datasets and recordings acquired with different stethoscope hardware and using wearable multi-sensor arrays, to assess robustness to domain shift and recording variability.
From a system perspective, future work will address the integration of a digital stethoscope front-end and the design of a custom embedded board. This would overcome the main limitation of the current prototype, which does not directly acquire heart sound recordings, and would make the system fully autonomous and suitable for real-world deployment.
Finally, the detailed profiling of the processing pipeline enables the evaluation of alternative MCU platforms, including devices with lower computational and memory resources, to identify the most suitable hardware configuration for the final application.

% Furthermore, explainability techniques, such as saliency or relevance mapping, could be combined with uncertainty quantification to provide clinicians and non-expert users with interpretable visual feedback.
  
%\section*{Acknowledgment}
%\textcolor{red}{Add here the DET funding? %Or in the unnumbered footnote on the %first page?}
%$textcolor{red}{\textbf{TO DO}} The preferred spelling of the word ``acknowledgment'' in American English is 
%without an ``e'' after the ``g.'' Use the singular heading even if you have 
%many acknowledgments. Avoid expressions such as ``One of us (S.B.A.) would 
%like to thank $\ldots$ .'' Instead, write ``F. A. Author thanks $\ldots$ .'' In most 
%cases, sponsor and financial support acknowledgments are placed in the 
%unnumbered footnote on the first page, not here.
\section*{References}
\bibliographystyle{IEEEtran}
\bibliography{bibliography}
%\begin{IEEEbiography}[{\includegraphics[width=1in,height=1.25in,clip,keepaspectratio]{a1.png}}]{First A. Author} (Fellow, IEEE) and all authors may include 
%\end{IEEEbiography}
%\begin{IEEEbiographynophoto}{Second B. Author,} photograph and biography not available at the
%time of publication.
%\end{IEEEbiographynophoto}
%\begin{IEEEbiographynophoto}{Third C. Author Jr.} (Member, IEEE), photograph and biography not available at the
%time of publication.
%\end{IEEEbiographynophoto}
\end{document}